\documentclass[pra,twoside,showpacs,superscriptaddress,twocolumn]{revtex4}

\usepackage{latexsym}
\usepackage[dvips]{graphics}
\usepackage{psfig}

\begin{document}

\title{Experimental phase-covariant cloning of polarization states of single photons}

\author{Anton\'{\i}n \v{C}ernoch}
\affiliation{Department of Optics, Palack\'y University,
     17.~listopadu 50, 772\,00 Olomouc, Czech~Republic}

\author{Lucie Bart{\accent23 u}\v{s}kov\'{a}}
\affiliation{Department of Optics, Palack\'y University,
     17.~listopadu 50, 772\,00 Olomouc, Czech~Republic}

\author{Jan Soubusta}
\affiliation{Joint Laboratory of Optics of Palack\'{y} University and
     Institute of Physics of Academy of Sciences of the Czech Republic,
     17. listopadu 50A, 779\,07 Olomouc, Czech Republic}

\author{Miroslav Je\v{z}ek}
\affiliation{Department of Optics, Palack\'y University,
     17.~listopadu 50, 772\,00 Olomouc, Czech~Republic}

\author{Jarom{\'\i}r Fiur\'{a}\v{s}ek}
\affiliation{Department of Optics, Palack\'y University,
     17.~listopadu 50, 772\,00 Olomouc, Czech~Republic}

\author{Miloslav Du\v{s}ek}
\affiliation{Department of Optics, Palack\'y University,
     17.~listopadu 50, 772\,00 Olomouc, Czech~Republic}

\date{\today}

\begin{abstract}
The experimental realization of optimal symmetric
 phase-covariant $1\to2$ cloning of qubit
states is presented. The qubits are represented by
polarization states of photons generated by spontaneous
parametric down-conversion. The experiment is based on the interference of two
photons on a custom-made beam splitter with different splitting ratios
for vertical and horizontal polarization components. From
the measured data we have estimated the implemented cloning
transformation using the maximum-likelihood method. The
result shows that the realized transformation is very close
to the ideal one and  the map fidelity reaches 94\,\%. 

\end{abstract}


\maketitle




Unknown quantum states cannot be perfectly copied \cite{Wootters82}. The 
no-cloning theorem is a direct consequence of the superposition principle and linearity 
of quantum mechanics. The impossibility to duplicate 
an unknown  quantum state without introducing noise is exploited by the 
modern quantum communication protocols and lies at the heart of the security 
of quantum key distribution schemes \cite{Gisin02}. Although perfect copying 
is forbidden  one may nevertheless copy the states in an approximate way. 
The optimal quantum cloning machine introduced  by Bu\v{z}ek and Hillery in 1996
yields clones whose fidelity with respect to the input state is maximum
possible \cite{Buzek96}. Since then, the quantum cloning has been investigated 
by numerous authors, see recent reviews \cite{Scarani05,Cerf05} and references therein. 

During recent years growing attention has been devoted to the 
experimental implementation of the various cloning machines. Optimal universal
cloning  of polarization states of photons has been demonstrated by several
groups by exploiting the process of stimulated parametric down-conversion 
\cite{Linares02,Fasel02,DeMartini04} or the
bunching of photons which interfere on a beam splitter \cite{Ricci04,Irvine04,Khan04}. 
Universal cloning
machines producing three copies \cite{Masullo05} and asymmetric universal 
cloning machines \cite{Zhao05} were also reported.  
The universal machine copies equally well all states.  
In many situations, however, we need to copy only a subset
of the states. In particular, the phase-covariant quantum cloning machine 
\cite{Niu99,Bruss00} copies
equally well all states on the equator of the Poincar\'{e} sphere, i.e. all balanced
superpositions of the basis states $|0\rangle$ and $|1\rangle$,
$|\psi\rangle=\frac{1}{\sqrt{2}}(|0\rangle+ e^{i\phi}|1\rangle)$, where the
phase $\phi$ is arbitrary. Due to the restriction to a smaller set of states,
the $1\rightarrow 2$  phase-covariant cloner achieves slightly higher cloning
fidelity $F_{\mathrm{pc}}=\frac{1}{2}(1+\frac{1}{\sqrt{2}})\approx 0.854$ than the $1\rightarrow 2$ universal  cloner,
whose fidelity reads $F_{\mathrm{univ}}=\frac{5}{6} \approx 0.833$.

The optimal economical phase-covariant cloning transformation requires only a
single blank copy in addition to the input qubit 
to be cloned and reads \cite{Niu99},
\begin{eqnarray}
|0\rangle&\rightarrow & |00\rangle,  \nonumber \\
 |1\rangle &\rightarrow & \frac{1}{\sqrt{2}}(|01\rangle+|10\rangle). 
\label{pc_cloning}
\end{eqnarray}
The optimal phase-covariant cloner represents a very efficient individual
eavesdropping attack on the BB84 quantum key distribution protocol
\cite{Fuchs97,Cerf02}. By using the
asymmetric version of the cloning machine the eavesdropper can in an optimal way
choose the trade-off between the information she gains on the secret key 
and the amount of noise that is added to the state which is sent down the
communication line. It then comes as a surprise that, to the best of our
knowledge, the $1\rightarrow 2$  phase covariant cloning machine has not yet been
demonstrated experimentally for optical qubits. This machine has been realized
in an NMR experiment \cite{Du05}, which however is not suitable for quantum 
communication applications where cloning of the states of single photons 
is desirable.  Note also that  Sciarrino and De Martini implemented 
the $1\rightarrow 3$ phase-covariant 
cloning of photonic qubits \cite{Sciarrino05}.

Here we report on the experimental realization of the optimal phase-covariant cloning
transformation (\ref{pc_cloning}) for the polarization states of photons. Our experimental setup
follows the theoretical proposal put forward in Ref. \cite{Fiurasek03}. The  cloning is achieved
by an interference of a signal photon whose state should be cloned with an ancilla
photon prepared in a fixed polarization state on an particularly tailored 
unbalanced beam splitter. We measure the fidelities of the two clones for a wide
variety of input states and perform a maximum-likelihood estimation of the
cloning operation which provides a detailed characterization of our experimental
scheme. We find that due to the imperfections of our beam splitter the cloner is
unbalanced and the fidelities of the two clones slightly differ. We actively compensate for this
effect and symmetrize the cloner by inserting a tilted glass plate into the path 
of one photon.


Let us begin with a theoretical description of the cloning setup, see Fig. 1. In our scheme, the
qubits are represented by polarization states of single photons. We identify the
computational basis states $|0\rangle$ and $|1\rangle$ with the vertical $|V\rangle$
and horizontal $|H\rangle$ polarization states, respectively.  The ancilla photon is
initially vertically polarized while the signal photon can be prepared in an
arbitrary state.  The two photons interfere on an
unbalanced beam splitter BS which exhibits different real amplitude transmittances
$t_H$, $t_V$ and reflectances $r_H$, $r_V$ for the
horizontal and vertical polarizations. We use the notation 
$R_j=r_j^2$ and $T_j=t_j^2$ for the intensity reflectances and transmittances and we
have $R_j+T_j=1$ for a lossless beam splitter.
As shown in Ref. \cite{Fiurasek03}, a symmetric cloning
requires $r_H=t_V$ and $t_H=-r_V$. 


\begin{figure}[!t!]
 \centerline{ \resizebox{0.9\hsize}{!}{\rotatebox{0}{\includegraphics*{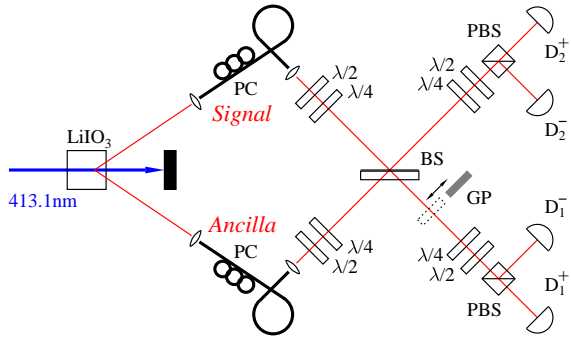}}}}
  \caption{Experimental setup. For details, see text.}
  \label{fig1}
\end{figure}


In the experiment, we accept only the events when
there is a single photon detected in each output port of the beam splitter. The
cloning transformation is thus implemented conditionally, similarly to other
optical cloning experiments.  The conditional transformation reads \cite{Fiurasek03}
\begin{eqnarray}
|V\rangle_S|V\rangle_A & \rightarrow & (r_V^2-t_V^2)|VV\rangle, \nonumber \\[2mm]
|H\rangle_S|V\rangle_A & \rightarrow & r_V t_V(|HV\rangle+|VH\rangle).
\label{HV_transformation}
\end{eqnarray}
If $r_V$   and $t_V=\sqrt{1-r_V^2}$ are chosen such that $\sqrt{2}r_Vt_V=r_V^2-t_V^2$
then the device implements the optimal phase-covariant cloning transformation
(\ref{pc_cloning}).
This happens when $R_V=\frac{1}{2}(1+\frac{1}{\sqrt{3}}) \approx 0.789$.
In order to implement the cloning operation we thus need an  unbalanced 
beam splitter with $79 \%$ reflectance for vertical polarization and $21\%$
reflectance for horizontal polarization. For this optimal beam splitter 
the probability of successful cloning  reads 
$P_{\mathrm{succ}}=2 R_V T_V=\frac{1}{3}$.


The scheme of our experimental setup is shown in Fig.~\ref{fig1}. A
krypton-ion cw laser (413.1\,nm, 120\,mW) is used to pump a
10-mm-long LiIO$_3$ nonlinear crystal (NLC) cut for
frequency-degenerate (826.2\,nm) type-I parametric
down-conversion. The pairs of photons generated by
spontaneous parametric down-conversion (SPDC) manifest
tight time correlations (i.e., very exact coincidences of
detection events). Both photons are coupled into
single-mode optical fibers that provide spatial-mode
filtration. The output beams from the fibers are set by the polarization 
controllers (PC) to have horizontal linear polarizations. Other polarization
states are prepared by means of half-wave plates and
quarter-wave plates ($\lambda/2$, $\lambda/4$) -- the
ancilla polarization state is fixed to a vertical linear
polarization, the signal polarization state is varied. 
The accuracy of polarization-angle settings was better than $\pm 1^\circ$.
Then the photons enter a custom-made unbalanced beam splitter (BS)
manufactured by Ekspla. The measured splitting ratios of the BS 
are 76:24 for vertical and 18:82 for horizontal polarization components,
close to the required optimal splitting ratios of 79:21 and 21:79, respectively.
 This special beam splitter is a key component of our setup.
 Finally there are two detection blocks that can detect two chosen
orthogonal polarizations. Each block consists of  quarter-
and half-wave plates, polarization beam splitter (PBS), and
two detectors. Detectors D$_{1}^{+}$, D$_{1}^{-}$, D$_{2}^{+}$,
D$_{2}^{-}$ are Perkin-Elmer single-photon counting modules
(employing silicon avalanche photodiodes with quantum
efficiency $\eta\approx 58\,\%$ and dark counts about
120\,s$^{-1}$). In each measurement the wave-plates are set such that the click of $D^{+}$ indicates
projection onto the input state of the signal photon while click of $D^{-}$
heralds projection onto the orthogonal state.
The signals from detectors are processed by
four-input coincidence module. Tiltable glass plate GP is
used to compensate the imperfection of the beam splitter by
implementing polarization dependent losses in one output
arm. This is necessary to implement \emph{symmetric}
cloning transformation.

\begin{figure}[!t!]
\centerline{\resizebox{0.9\hsize}{!}{\includegraphics*{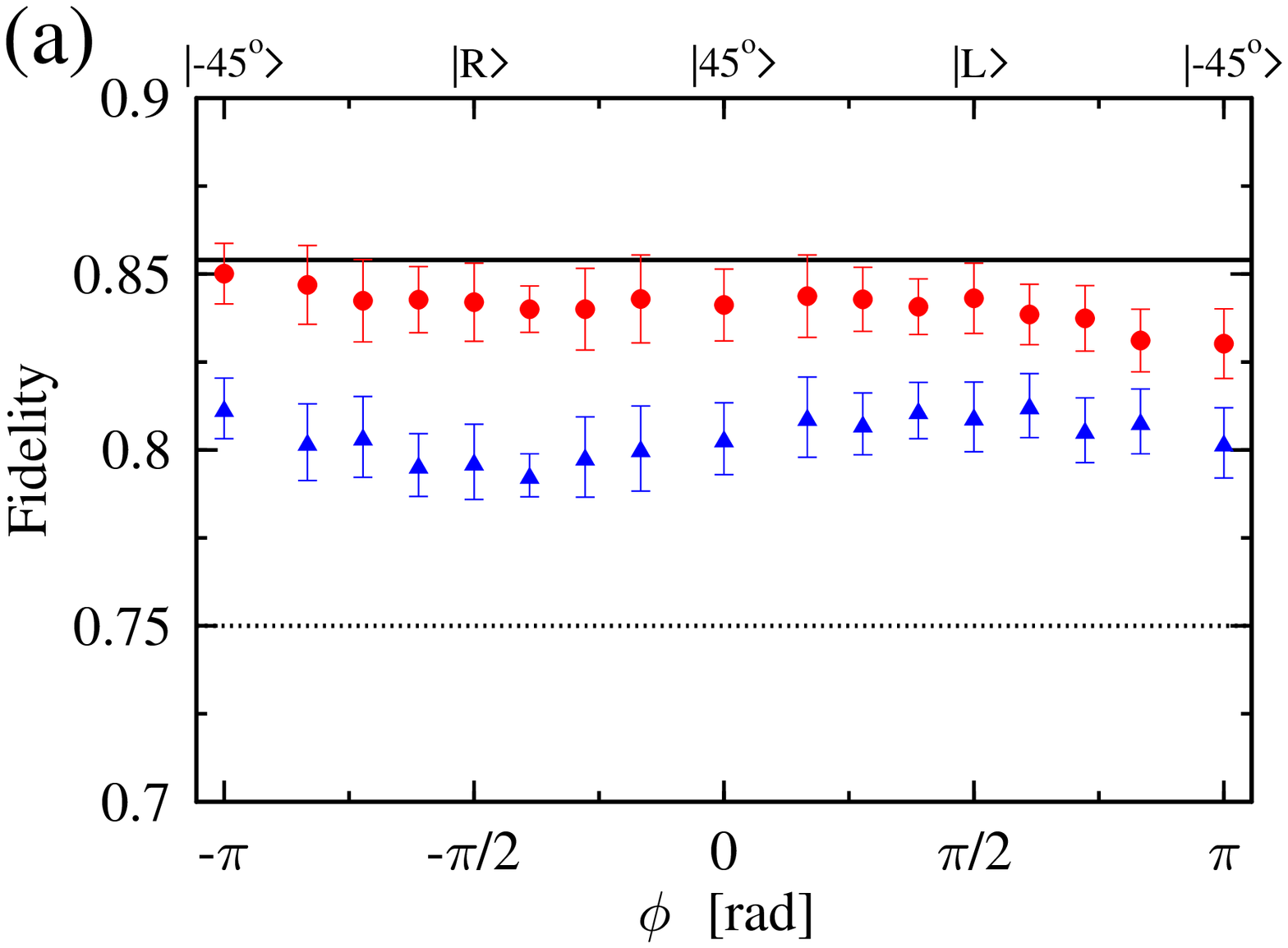}}}
\centerline{\resizebox{0.9\hsize}{!}{\includegraphics*{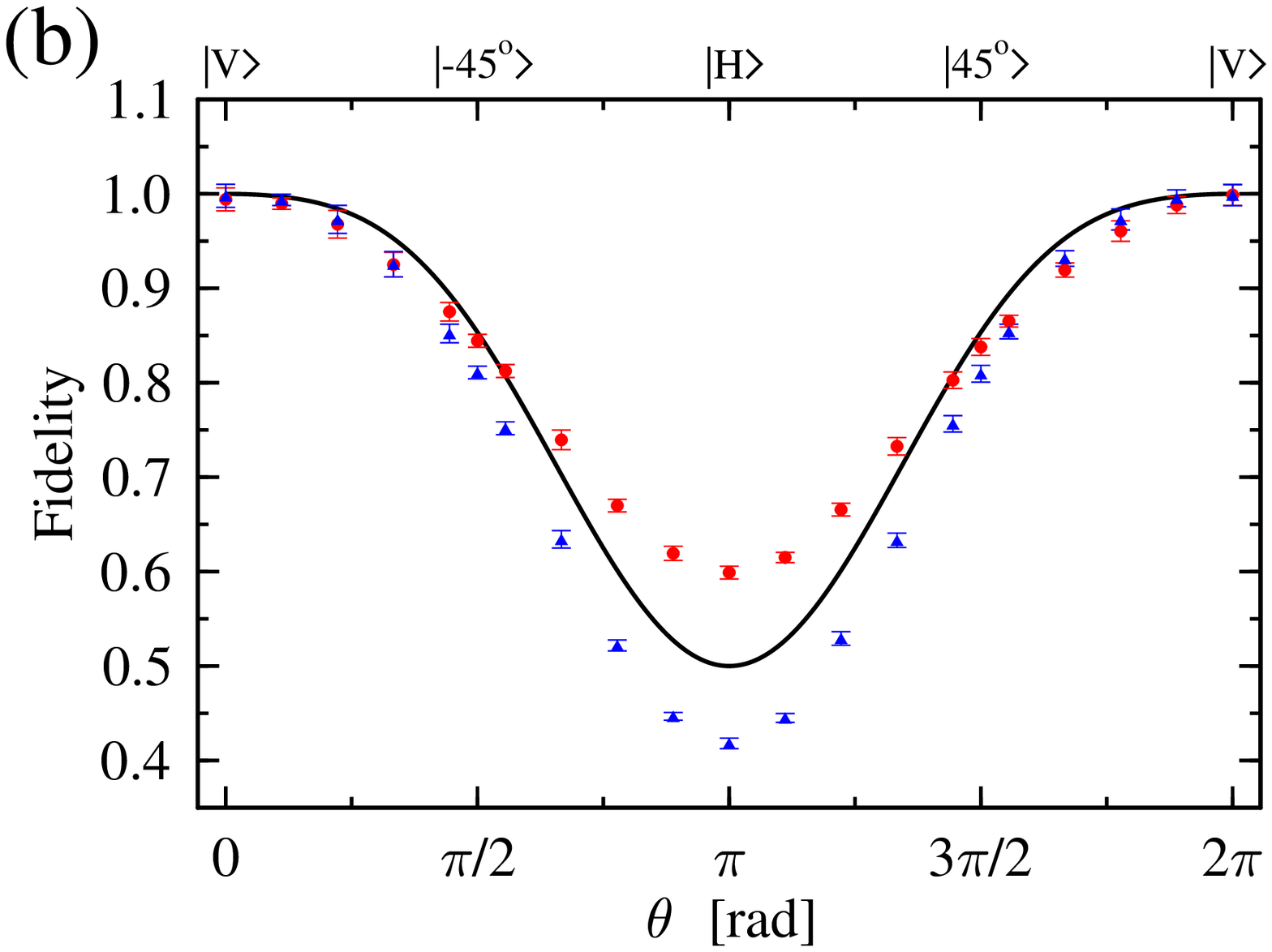}}}
  \caption{Fidelities $F_1$ (red circles) and $F_2$ (blue triangles)
   as functions of the input-state parameters. (a) $\phi$ is varied while 
   $\theta = \pi/2$ is fixed, (b) $\phi=0$ is fixed and $\theta$ is varied.
   Symbols denote experimental data, lines represent theoretical predictions.}
  \label{fig2}
\end{figure}


The cloning is successful if each photon goes by different
output arm. Therefore, we measure coincidences between the
detectors at two different outputs $C^{++}, C^{+-}, C^{-+},
C^{--}$ where the first sign concerns the lower arm and the
second one the upper arm, ``$+$'' means the correct result
(the same as the input state), ``$-$'' means the wrong one.
Fidelity of the first (second) clone thus reads
\begin{eqnarray}
  F_1 = \frac{C^{++} + C^{+-}}{C^{++} + C^{+-} + C^{-+} + C^{--}},
  \nonumber \\
  F_2 = \frac{C^{++} + C^{-+}}{C^{++} + C^{+-} + C^{-+} + C^{--}}.
 \label{Fidel}
\end{eqnarray}
Probability of successful cloning can be determined as
\begin{equation}
  P_{\mathrm{succ}} = \frac{C^{++} + C^{+-} + C^{-+} + C^{--}}{C_{\mathrm{tot}}},
 \label{Probab}
\end{equation}
where the total rate of events $C_{\mathrm{tot}}$ is
obtained from the sum of all coincidence events $C_{\mathrm{sum,dis}}$ measured
with mutually delayed (i.e., distinguishable) input
photons. For this measurement the signal photon is prepared in the $-45^{\circ}$ linear 
polarization and the ancilla remains vertically linearly polarized.
The delay is realized by prolonging one input
arm by $120\,\mu$m. A simple calculation reveals that
$C_{\mathrm{tot}}=C_{\mathrm{sum,dis}}/Q$ where $Q=(T_V^2+R_V^2+T_VT_H+R_VR_H)/2$
is the probability that there would be a single photon in each output port of the BS. 
Numerically, we get $Q=0.484$.

We prepared various input polarization states 
$|\psi\rangle=\cos \frac{\theta}{2}|V\rangle+\sin\frac{\theta}{2}e^{i\phi}|H\rangle$
and  measured fidelities $F_1, F_2$  as functions of
$\phi$ and $\theta$. First we investigated cloning of states on the equator of the
Poincar\'{e} sphere. We fixed $\theta = \pi/2$ and varied $\phi$. The
results are shown in Fig.~\ref{fig2}(a). Each data point at presented plots has 
been derived from ten 5-second measurement periods. Symbols denote
experimental data, lines represent theoretical predictions. The upper line indicates
the fidelity of the optimal symmetric phase-covariant cloner $F_{\mathrm{pc}}\approx 0.854$
while the lower line shows the fidelity of the optimal semiclassical cloning strategy
based on the optimal estimation of the state \cite{Derka98} followed by preparation of two copies,
$F_{\mathrm{est}}=0.75$. 

We can see that the fidelities of the two clones differ by approximately $4\%$.
This asymmetry can be attributed to the beam splitter whose reflectances somewhat
differ from the ideal ones, as discussed above. The mean fidelities of the first and
second clone averaged over the equator of the Poincar\'{e} sphere read $F_1=84.1  \pm
0.2 \, \%$ and $F_2=80.4 \pm 0.2\, \%$. Both $F_1$ and $F_2$ are below the theoretical maximum
$F_{\mathrm{pc}}$ for symmetric cloner.  This is due to several experimental imperfections, the
most important ones being the imperfect overlap of the two photons on a beam splitter
resulting in reduced visibility of the Hong-Ou-Mandel interference \cite{Hong87}, the imperfect
setting of the waveplates, and the random coincidences because of relatively
wide coincidence window ($20$ ns). 
The fidelities are almost constant and independent of $\phi$ which confirms 
the phase covariance of the cloning device. The small modulation 
of $F_1$ and $F_2$ is probably 
caused by an imperfect preparation of the ancilla state whose deviation 
from $|V\rangle$ would result in the observed oscillation of fidelities.

\begin{figure}
\centerline{\psfig{figure=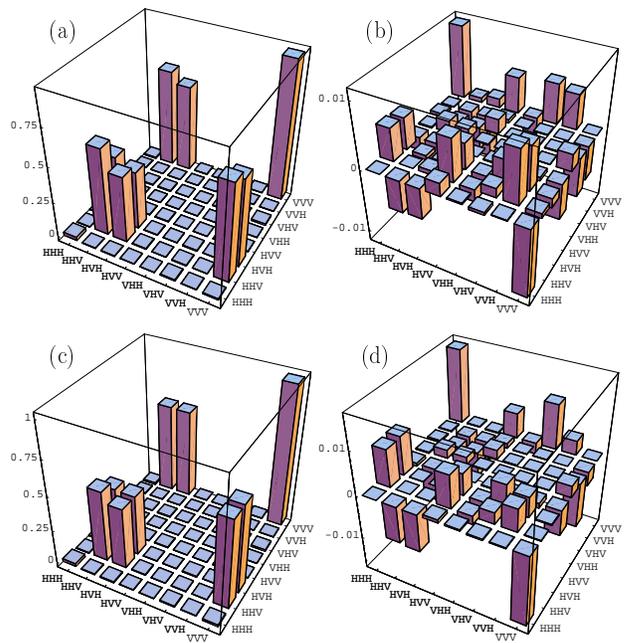,width=0.95\linewidth}}
\caption{Completely positive cloning map $E_{\mathrm{cl}}$
estimated from the experimental data. Panels (a) and (b) show 
the real and imaginary parts of $E_{\mathrm{cl}}$ before compensation and panels (c) and (d) show the same for
the map after compensation with a tilted glass plate. To facilitate the comparison,
in both cases the map is normalized to $\mathrm{Tr}[E_{\mathrm{cl}}]=2$.}
\label{fig3}
\end{figure}

The transformation (\ref{pc_cloning}) actually optimally clones all states 
on the northern hemisphere of the Poincar\'{e} sphere \cite{Fiurasek03}, 
i.e. all states with $|\theta| \leq \frac{\pi}{2}$.
We have studied the cloning of  states with various $\theta$ and the results 
for $\phi=0$ are shown in Fig. \ref{fig2}(b). We can see that for 
$\theta < \frac{\pi}{2}$ and $\theta > \frac{3\pi}{2}$ the observed fidelities are in good agreement with the
theoretical values indicated by a solid line. Note that we have carried out measurements also 
for $\frac{\pi}{2}<\theta <\frac{3\pi}{2}$. Although the cloning machine ceases to be optimal in
this region, the measured results provide a valuable characterization of the cloning
machine. In particular, the asymmetry of the cloner  is most clearly visible 
for the input state $|H\rangle$ where the two fidelities differ most significantly. 
We have also performed measurements similar to those showed in Fig. \ref{fig2}(b) but with
$\phi=\frac{\pi}{2}$. The results are very similar and are not shown here. 

The cloning transformation (\ref{pc_cloning}) is an isometry, i.e. 
a deterministic operation. The success probability of a conditional
implementation of such operation should not depend on the input state. 
In our case this means that the total number of 
coincidences $C_{\mathrm{sum}}=C^{++}+C^{+-}+C^{-+}+C^{--}$ should
be constant. In the experiment we observe that $C_{\mathrm{sum}}$ remains practically 
constant  as we vary $\phi$. The maximal relative change 
of $C_{\mathrm{sum}}$ when $\theta$  is varied from $0$ to $\pi$ is about $8\%$. 
This confirms that the transformation 
realized by our scheme is close to a deterministic operation (albeit implemented
conditionally). The changes in $C_{\mathrm{sum}}$ can be attributed to the difference 
of the actual splitting ratios of the BS from the ideal ones.

In order to characterize the cloning transformation more  completely, we
employed the quantum process tomography. Using all collected experimental data
we have  performed  a maximum likelihood (ML) estimation of the 
completely positive map $\mathcal{E}_{\mathrm{cl}}$ which fully specifies 
the cloning operation. According to  Jamiolkowski-Choi
isomorphism \cite{Jamiolkowski72}, any completely positive map $\mathcal{E}$ is isomorphic to 
a positive semidefinite operator $E$ on the tensor product of input 
and output Hilbert spaces. For any (generally mixed) input state $\rho_{\mathrm{in}}$
the corresponding output state $\rho_{\mathrm{out}}=\mathcal{E}(\rho_{\mathrm{in}})$ 
can be determined as
$\rho_{\mathrm{out}}=\mathrm{Tr}_{\mathrm{in}}[\rho_{\mathrm{in}}^T\otimes
\openone_{\mathrm{out}} \, E]$,  where $T$ indicates transposition with respect to a
fixed basis and $\openone_{\mathrm{out}}$ denotes an identity operator 
on $\mathcal{H}_{\mathrm{out}}$.

Here the input and output Hilbert spaces are
Hilbert spaces of one and two qubits, respectively, hence $E_{\mathrm{cl}}$ is an $8 \times 8$
Hermitian positive semidefinite matrix.  The ML estimation yields a
transformation that is most likely to produce the observed
experimental data. The advantage of this nonlinear statistical estimation method is
that it guarantees the complete positivity of the estimated operation. In our case
the map is not exactly trace preserving, so we have to estimate a general 
trace-decreasing completely positive map. We follow the procedure outlined in Ref.
\cite{Hradil04}. 
We extend the output Hilbert space to include a fifth virtual sink level
$|S \rangle$. The rate of events associated with the detection of the state
$|S \rangle$ is set to $C^{S }=C_{\mathrm{tot}}-C_{\mathrm{sum}}$.
 On this extended output
Hilbert space we reconstruct a trace-preserving operation using the well established
iterative algorithm \cite{Hradil04,Jezek03}. From the resulting map represented by a $10 \times 10$
matrix we extract the $8\times 8$ sub-matrix which characterizes the 
(generally trace decreasing) cloning operation.

The results are shown in Fig. 3(a,b). The map which corresponds to the optimal cloning
operation (\ref{pc_cloning}) reads 
$E_{\mathrm{opt}}=|E_{\mathrm{opt}}\rangle\langle E_{\mathrm{opt}} |$, where
\begin{equation}
|E_{\mathrm{opt}}\rangle= \frac{1}{\sqrt{2}}|H\rangle
(|HV\rangle+|VH\rangle)+|V\rangle|VV\rangle.
\label{E_cloning}
\end{equation}
The similarity of the estimated map $E_{\mathrm{cl}}$ with the optimal 
map $E_{\mathrm{opt}}$
can be quantified by the map fidelity, defined as
\begin{equation}
\mathcal{F}= \frac{\langle
E_{\mathrm{opt}}|E_{\mathrm{cl}}|E_{\mathrm{opt}}\rangle}{2
\mathrm{Tr}[E_{\mathrm{cl}}]}.
\label{Fidelity_map}
\end{equation}
For the map shown in Fig. 3(a,b) we obtain $\mathcal{F}=93\%$.

\begin{figure}
\centerline{\resizebox{0.9\hsize}{!}{\includegraphics*{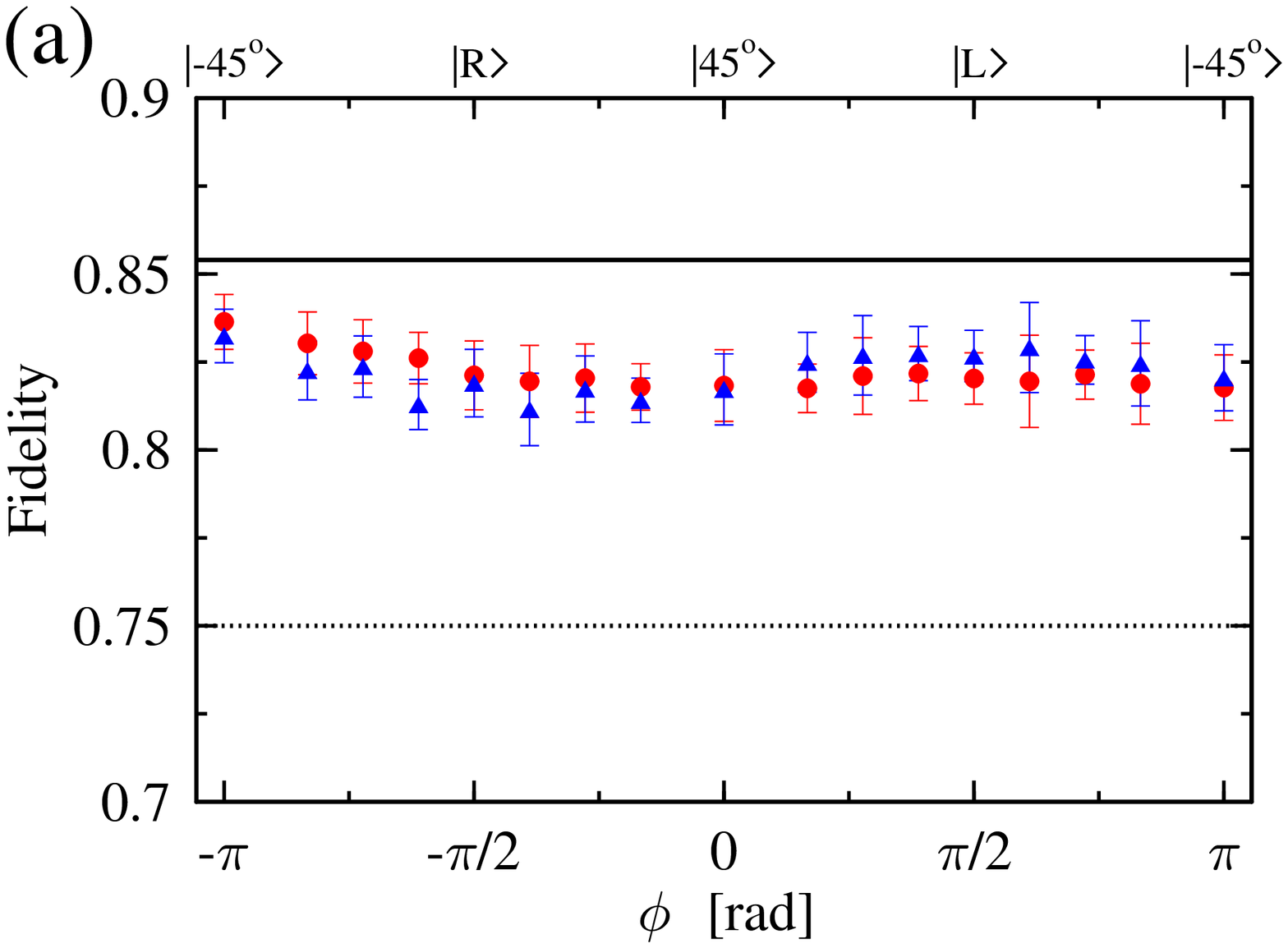}}}
\centerline{\resizebox{0.9\hsize}{!}{\includegraphics*{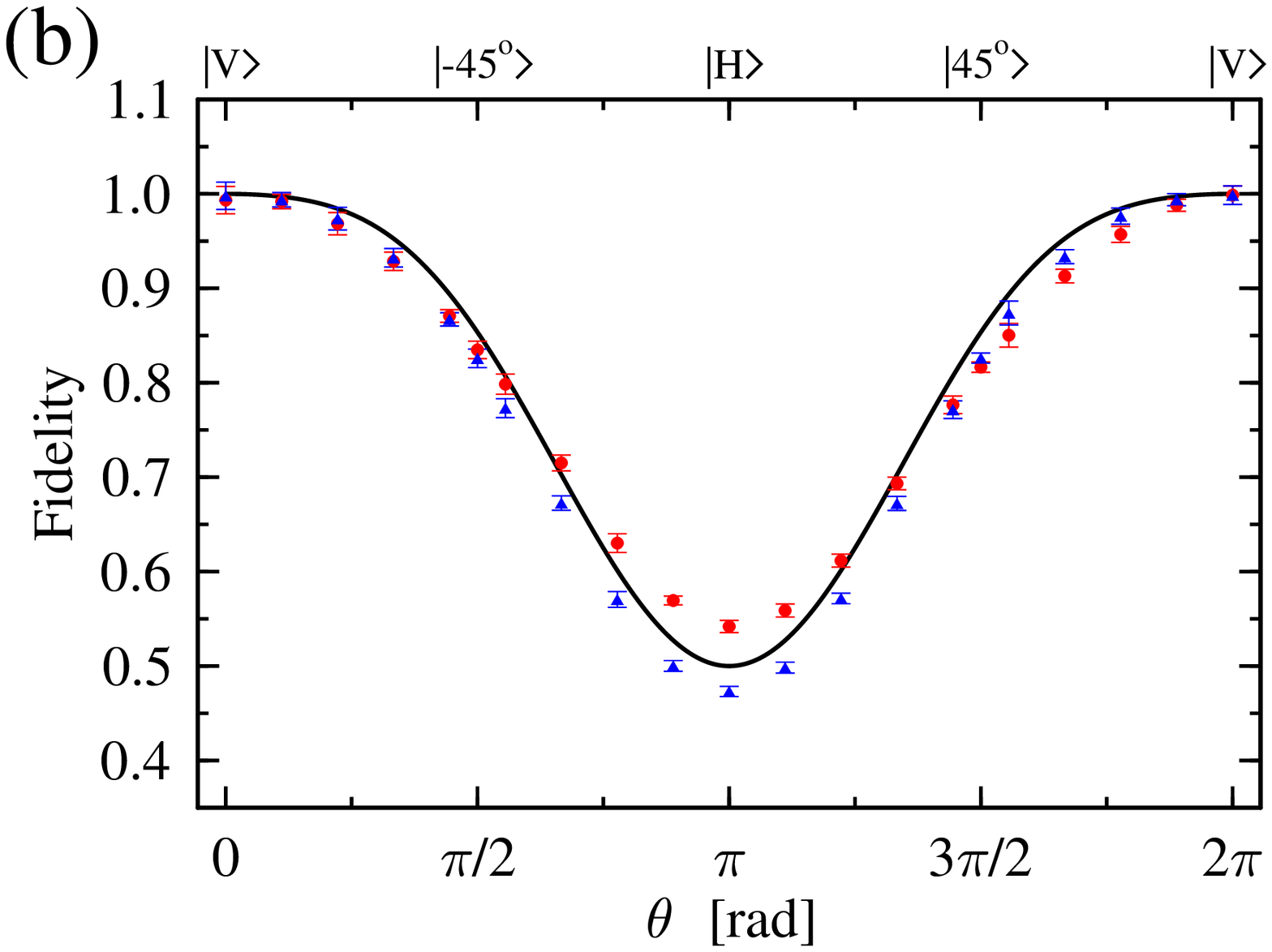}}}
  \caption{The same as Fig. 2 but with glass plate filter inserted in the setup.}
  \label{fig4}
\end{figure}

The asymmetry between the two clones is clearly revealed in the reconstructed 
map as the difference between the $HHV$ and $HVH$ matrix elements. As noted above,
this is caused by the imperfections of the beam splitter, whose transmittances and
reflectances do not precisely satisfy the symmetry condition $r_H=t_V$ and
$t_H=-r_V$. The mapping accomplished by such general beam splitter can be expressed
as
\begin{eqnarray}
|VV\rangle & \rightarrow & (R_V-T_V) |VV \rangle, \nonumber \\
 |HV \rangle & \rightarrow & r_H r_V |VH \rangle-t_H t_V |HV\rangle. 
 \label{unbalanced_BS}
\end{eqnarray}
If $|r_V r_H|\neq |t_V t_H|$ then the cloner is asymmetric and produces two copies
with different fidelities. To recover a symmetric copying machine, we apply 
an active filtering operation on one of the clones. A tilted glass plate (GP)
is inserted into the path of the photon in one output port of the BS. This plate
acts as a filter with different transmittances $\eta_H$ and $\eta_V$ for vertical
and horizontal polarizations. The ratio $\eta_H/\eta_V$  can be controlled by
changing the tilt angle of the plate. With the plate present the transformation
(\ref{unbalanced_BS}) changes to
\begin{eqnarray}
|VV\rangle &\rightarrow &\eta_V (R_V-T_V) |VV \rangle, \nonumber \\
 |HV \rangle &\rightarrow &\eta_V r_H r_V |VH \rangle-\eta_H t_H t_V |HV\rangle. 
 \label{filter}
\end{eqnarray}
If we position the plate such that $\eta_H/\eta_V=|r_H r_V|/|t_H t_V|$
then we recover a symmetric cloning transformation.

The fidelities measured with the GP inserted in the setup are shown in Fig.
\ref{fig4}. We can see that the cloner is successfully symmetrized and 
the fidelities of the
two clones of the equatorial qubits are practically identical. The mean fidelities
coincide within the measurement error, $F_1=F_2=82.2 \pm 0.2 \, \%$. 
The symmetrization is also clearly witnessed by Fig. \ref{fig4}(b) where we can see that 
the difference between the two fidelities for input state $|H\rangle$ is much less
than before compensation, c.f. Fig. \ref{fig2}(b). The filtering reduced 
the relative variation of the total number of coincidences $C_{\mathrm{sum}}$ 
over all input states from $8\%$ to $6\%$ thus making the transformation 
very close to a (conditionally implemented) deterministic operation. 
The measured average success  probability 
$P_{\mathrm{succ}}=0.292 \pm 0.005$ is in a good agreement with the theoretical 
value $1/3$. We have again carried out a reconstruction of the cloning 
map and the results are shown in Fig. \ref{fig3}(c,d). The filtering increased
the fidelity of the map with respect to the optimal transformation 
$E_{\mathrm{opt}}$ and we find $\mathcal{F}=94\%$.

In summary, we have experimentally implemented the optimal phase-covariant cloning 
of polarization states of single photons. The imperfections of the specifically
tailored beam splitter which forms the core part of the cloning setup were
compensated by a glass plate filter. In the future work we plan to investigate the
possibility of optimal asymmetric phase-covariant cloning by using properly tilted
glass plate filters. Another goal is to improve the parameters of the experimental
setup such as to achieve for equatorial qubits cloning fidelities higher than
the fidelity of the optimal universal cloning machine.

This research was supported by the projects LC06007 and MSM6198959213
of the Ministry of Education of the Czech Republic and by the SECOQC project 
of the EC (IST-2002-506813). M. Je{\v{z}}ek acknowledges financial support from
Grant \#202/05/0498 of GACR. We would like to thank Ekspla company for 
design and manufacturing  the custom unbalanced beam splitter and Romas Remeika 
from  Ekspla for his long-standing support and many discussions on
specifically tailored optical components.


\begin{thebibliography}{99}



\bibitem{Wootters82}
W.K. Wootters and W.H. Zurek, Nature (London) \textbf{299}, 802 (1982);
D. Dieks, Phys. Lett. \textbf{92A}, 271 (1982).


\bibitem{Gisin02}
N. Gisin, G. Ribordy, W. Tittel, and H. Zbinden,
Rev. Mod. Phys. \textbf{74}, 145 (2002).



\bibitem{Buzek96}
V. Bu\v{z}ek and M. Hillery, Phys. Rev. A \textbf{54}, 1844 (1996).

\bibitem{Scarani05}
V. Scarani, S. Iblisdir, N. Gisin, and A. Ac\'{\i}n,
Rev. Mod. Phys. \textbf{77}, 1225 (2005).

\bibitem{Cerf05}
N.J. Cerf and J. Fiur\'{a}\v{s}ek, quant-ph/0512172.




\bibitem{Linares02}
A. Lamas-Linares,  C. Simon, J.C. Howell, and D. Bouwmeester, 
Science \textbf{ 296}, 712 (2002).


\bibitem{Fasel02}
S. Fasel, N. Gisin, G. Ribordy, V. Scarani, and H. Zbinden,
Phys. Rev. Lett. \textbf{ 89}, 107901 (2002).

\bibitem{DeMartini04}
F. De Martini, D. Pelliccia, and F. Sciarrino,
Phys. Rev. Lett. \textbf{92}, 067901 (2004).

\bibitem{Ricci04}
M. Ricci, F. Sciarrino, C. Sias, and F. De Martini
Phys. Rev. Lett. 92, 047901 (2004).

\bibitem{Irvine04} 
W.T. M. Irvine, A. Lamas-Linares, M. J. A. de Dood, and D. Bouwmeester,
 Phys. Rev. Lett. \textbf{92}, 047902 (2004). 

\bibitem{Khan04}
I. A. Khan and J. C. Howell, Phys. Rev. A \textbf{70}, 010303 (2004). 


\bibitem{Masullo05}
L. Masullo, M. Ricci, and F. De Martini,
Phys. Rev. A \textbf{72}, 060304 (2005).  

\bibitem{Zhao05}
Z. Zhao, A.-N. Zhang, X.-Q. Zhou, Y.-A. Chen, C.-Y. Lu, A. Karlsson, and J.-W. Pan,
Phys. Rev. Lett. \textbf{95}, 030502 (2005). 



\bibitem{Niu99}
C.-S. Niu and R.B. Griffiths, Phys. Rev. A \textbf{ 60}, 2764 (1999).

\bibitem{Bruss00}
D. Bruss M. Cinchetti, G. M. D'Ariano, and C. Macchiavello, 
Phys. Rev. A \textbf{62}, 012302 (2000).



\bibitem{Fuchs97}
C.A. Fuchs,  N. Gisin, R. B. Griffiths, C.-S. Niu, and A. Peres,
Phys. Rev. A \textbf{56}, 1163 (1997).


\bibitem{Cerf02}
N.J. Cerf, M. Bourennane, A. Karlsson, and N. Gisin, 
Phys. Rev. Lett. \textbf{ 88}, 127902 (2002).

\bibitem{Du05}
J.-F. Du, T. Durt, P. Zou, H. Li, L. C. Kwek, C. H. Lai, C. H. Oh, and A. Ekert,
Phys. Rev. Lett. \textbf{94}, 040505 (2005). 

\bibitem{Sciarrino05}
F. Sciarrino and F. De Martini,
Phys. Rev. A \textbf{72}, 062313 (2005). 

\bibitem{Fiurasek03}
J. Fiur\'{a}\v{s}ek,
Phys. Rev. A \textbf{67}, 052314 (2003).



\bibitem{Derka98}
R. Derka, V. Bu\v{z}ek, and A. K. Ekert,
Phys. Rev. Lett. \textbf{80}, 1571 (1998).



\bibitem{Hong87}
C.K. Hong, Z.Y. Ou, and L. Mandel, 
Phys. Rev. Lett. \textbf{59}, 2044 (1987). 



\bibitem{Jamiolkowski72}
A. Jamiolkowski, Rep. Math. Phys. \textbf{ 3}, 275 (1972);
M.-D. Choi, Lin. Alg. Appl. \textbf{ 10}, 285 (1975).



\bibitem{Hradil04}
Z. Hradil, J. \v{R}eh\'{a}\v{c}ek, J. Fiur\'{a}\v{s}ek, and M. Je\v{z}ek, 
in \emph{Quantum states estimation}, 
M. G. A. Paris and J. \v{R}eh\'{a}\v{c}ek Eds., Lect. Not. Phys. 649
(Springer, Heidelberg, 2004).

\bibitem{Jezek03}
M. Je\v{z}ek, J. Fiur\'{a}\v{s}ek, and Z. Hradil, 
Phys. Rev. A \textbf{68}, 012305 (2003).  


\end{thebibliography}
\end{document}